\documentclass{elsart}

\usepackage{graphicx}
\usepackage{epsfig}
\usepackage{units}
\usepackage{amsmath}
\usepackage{amssymb}
\usepackage[figuresright]{rotating}
\newcommand{\hpt}{high $p_T$}

\begin{document}
\begin{frontmatter}
\title{\bf Quark Gluon Plasma\\ and Color Glass Condensate at RHIC?\\
The perspective from the BRAHMS experiment.}

\author[bucharest]{I.~Arsene},
\author[nbi]{I.~G.~Bearden},
\author[bnl]{D.~Beavis},
\author[bucharest]{C.~Besliu},
\author[nyc]{B.~Budick},
\author[nbi]{ H.~B{\o}ggild},
\author[bnl]{C.~Chasman},
\author[nbi]{C.~H.~Christensen},
\author[nbi]{P.~Christiansen},
\author[krakunc]{J.~Cibor},
\author[bnl]{R.~Debbe},
\author[oslo]{E.~Enger},
\author[nbi]{J.~J.~Gaardh{\o}je},
\author[nbi]{M.~Germinario},
\author[nbi]{O.~Hansen},
\author[nbi]{A.~Holm},
\author[oslo]{A.~K.~Holme},
\author[texas]{K.~Hagel},
\author[bnl]{H.~Ito},
\author[nbi]{E.~Jakobsen},
\author[bucharest]{A.~Jipa},
\author[ires]{F.~Jundt},
\author[bergen]{J.~I.~J{\o}rdre},
\author[nbi]{C.~E.~J{\o}rgensen},
\author[krakow]{R.~Karabowicz},
\author[bnl,kansas]{E.~J.~Kim},
\author[krakow]{T.~Kozik},
\author[nbi,oslo]{T.~M.~Larsen},
\author[bnl]{J.~H.~Lee},
\author[baltimore]{Y.~K.~Lee},
\author[oslo]{S.~Lindahl}
\author[oslo]{G.~L{\o}vh{\o}iden},
\author[krakow]{Z.~Majka},
\author[texas]{A.~Makeev},
\author[oslo]{M.~Mikelsen},
\author[texas,kansas]{M.~J.~Murray},
\author[texas]{J.~Natowitz},
\author[kansas]{B.~Neumann},
\author[nbi]{B.~S.~Nielsen},
\author[nbi]{D.~Ouerdane},
\author[krakow]{R.~P\l aneta},
\author[ires]{F.~Rami},
\author[nbi,bucharest]{C.~Ristea},
\author[bucharest]{ O.~Ristea},
\author[bergen]{D.~R{\"o}hrich},
\author[oslo]{B.~H.~Samset},
\author[nbi]{D.~Sandberg},
\author[kansas]{S.~J.~Sanders},
\author[bnl]{R.~A.~Scheetz},
\author[nbi]{P.~Staszel},
\author[oslo]{T.~S.~Tveter},
\author[bnl]{F.~Videb{\ae}k},
\author[texas]{R.~Wada},
\author[bergen]{Z.~Yin},
\author[bucharest]{I.~S.~Zgura},
\address[bnl]{Brookhaven National Laboratory, Upton, New York 11973, USA}
\address[ires]{Institut de Recherches Subatomiques et Universit{\'e} Louis
  Pasteur, Strasbourg, France}
\address[krakunc]{Institute of Nuclear Physics, Krakow, Poland}
\address[baltimore]{Johns Hopkins  University, Baltimore 21218, USA}
\address[krakow]{M. Smoluchkowski Inst. of Physics,
  Jagiellonian University, Krakow, Poland}
\address[nyc]{New York University, New  York 10003, USA}
\address[nbi]{Niels Bohr Institute,  University of Copenhagen, Copenhagen 2100, Denmark}
\address[texas]{Texas A$\&$M University, College Station, Texas, 17843, USA}
\address[bergen]{University of Bergen, Department of Physics, Bergen,  Norway}
\address[bucharest]{University of Bucharest, Romania}
\address[kansas]{University of Kansas, Lawrence, Kansas 66045, USA }
\address[oslo]{University of Oslo, Department of Physics, Oslo, Norway}

\begin{abstract}
  We review the main results obtained by the BRAHMS collaboration on
  the properties of hot and dense hadronic and partonic
  matter produced in ultrarelativistic heavy ion collisions at RHIC.
  A particular focus
  of this paper is to discuss to what extent the
  results collected so far by BRAHMS, and by the other three
  experiments at RHIC, can be taken as evidence for the
  formation of a state of deconfined partonic matter, the so called
  quark-gluon-plasma (QGP). We also discuss evidence for a possible precursor
  state to the QGP, i.e.  the proposed Color Glass
  Condensate.
\end{abstract}

\end{frontmatter}

\section{Introduction}

From the onset of the formulation of the quark model and the first
understanding of the nature of the binding and confining potential
between quarks about 30 years ago it has been conjectured that a
state of matter characterized by a large density of quarks and
gluons (together called partons) might be created for a fleeting
moment in violent nuclear collisions~\cite{GandMcL4_5}. This high
energy density state would be characterized by a strongly reduced
interaction between its constituents, the partons, such that these
would exist in a nearly free state. Aptly, this proposed state of
matter has been designated the quark gluon plasma (QGP). It is now
generally thought that the early universe was initially in a QGP
state until its energy density had decreased sufficiently, as a
result of the expansion of the universe, that it could make the
transition to ordinary (confined) matter.

Experimental attempts to create the QGP in the laboratory and
measure its properties have been carried out for more than 20
years, by studying collisions of heavy nuclei and analyzing the
fragments and produced particles emanating from such collisions.
During that period, center of mass energies per pair of colliding
nucleons have risen steadily from the $\sqrt{s_{NN}}\approx
\unit[1]{GeV}$ domain of the Bevalac at LBNL, to energies of
$\sqrt{s_{NN}}=\unit[5]{GeV}$ at the AGS at BNL, and to
$\sqrt{s_{NN}}=\unit[17]{GeV}$ at the SPS accelerator at CERN. No
decisive proof of QGP formation was found in the experiments at
those energies, although a number of signals suggesting the
formation of a "very dense state of matter" were found at the
SPS~\cite{SPSreview,SPSreview1}.

With the Relativistic Heavy Ion Collider, RHIC, at Brookhaven
National Laboratory, the center of mass energy in central
collisions between gold nuclei at
$\unit[100]{AGeV}+\unit[100]{AGeV}$ is almost $\unit[40]{TeV}$,
the largest so far achieved in nucleus-nucleus collisions under
laboratory conditions. This energy is so large that conversion of
a sizeable fraction of the initial kinetic energy into matter
production creates many thousands of particles in a limited volume
leading to unprecedented large energy densities and thus
presumably ideal conditions for the formation of the quark gluon
plasma.

RHIC started regular beam operations in the summer of year 2000
with a short commissioning run colliding Au nuclei at
$\sqrt{s_{NN}}=\unit[130]{GeV}$. The first full run at the top
energy ($\sqrt{s_{NN}}=\unit[200]{GeV}$) took place in the
fall/winter of 2001/2002.  The third RHIC run during the
winter/spring of 2003 focussed on d+Au and p+p reactions.
Recently, in 2004, a long high luminosity Au+Au run at
$\sqrt{s_{NN}}=\unit[200]{GeV}$ and a short run at
$\sqrt{s_{NN}}=\unit[62.4]{GeV}$  have been completed. The
collected data from the most recent runs are currently being
analyzed and only a few early results are thus available at the
time of writing of this document.

The aim here is to review the available information obtained from
the first RHIC experiments with the purpose of determining what
the experimental results, accumulated so far, allow us to say
about the high energy density matter that is created at RHIC in
collisions between heavy atomic nuclei.

We concentrate primarily on results from the BRAHMS detector, one
of the four detectors at RHIC, but naturally also refer to results
obtained by the other three experiments (STAR, PHENIX and PHOBOS)
insofar as they complement or supplement information obtained from
BRAHMS. The BRAHMS experiment is a two arm magnetic spectrometer
with excellent momentum resolution and hadron identification
capabilities. The two spectrometers subtend only a small solid
angle (a few msr) each, but they can rotate in the horizontal
plane about the collision point to collect data on hadron
production over a wide rapidity range (0-4), a unique capability
among the RHIC experiments. For details about the BRAHMS detector
system we refer the reader to ~\cite{BRAHMSNIM,BRAHMSmult}. The
large number of articles already produced by the four experiments
at RHIC may be found on their respective
homepages~\cite{RHICexpWWW}. Recent extensive theoretical reviews
and commentaries may be found in refs.
~\cite{intro_GandMcL,intro_shuryak,intro_jacobs}.

\section{What is the QGP and what does it take to see it?}

The predicted transition from ordinary nuclear matter, which
consists of hadrons inside which quarks and gluons are confined,
to the QGP, a state of matter in which quarks and gluons are no
longer confined to volumes of hadronic dimensions, can in the
simplest approach, be likened to the transition between two
thermodynamic states in a closed volume.

As energy is transferred to the lower energy state a phase
transition, akin to a melting or evaporation process, to the
higher energy state occurs. For a first order phase transition
(PT), the transformation of one state into the other occurs at a
specific temperature, termed the critical temperature, and the
process is characterized by absorption of latent heat during the
phase conversion, leading to a constancy or discontinuity of
certain thermodynamic variables as the energy density or
temperature is increased. In this picture, it is tacitly assumed
that the phase transition occurs between states in thermodynamic
equilibrium. From such thermodynamical considerations, and from
more elaborate models based on the fundamental theory for the
strong interaction, Quantum Chromo Dynamics (e.g. lattice QCD
calculations), estimates for the critical temperature and the
order of the transition can made. Calculations indicate that the
critical temperature should be $T_c \approx \unit[150-180]{MeV}$
in the case of a vanishing baryon chemical
potential~\cite{GandMcL_17_20} and the transition of second order.
In general, a decreasing critical temperature with increasing
chemical potential is expected. Likewise, at non-zero chemical
potential a mixed phase of coexisting hadron gas, HG, and QGP is
predicted to exist in a certain temperature interval around the
critical temperature. Recently calculational techniques have
progressed to the point of allowing an extension of the lattice
methods also to finite chemical potential. Such calculations also
suggest the existence of a critical point at larger chemical
potential above which, the transition may be of first order.

The transition from ordinary matter to the QGP is thus primarily a
deconfinement transition. However, it is also expected, due to the
vanishing interaction between partons in the QGP phase, that
hadron masses will be lowered. In the limit of chiral symmetry the
expectation value of the quark condensate ($< \bar{q} > $ vanishes
and opposite parity states (chiral partners) are degenerate. As a
consequence of the QGP to HG transition, the chiral symmetry is
broken and the hadrons acquire definite and nondegenerate masses.
According to lattice QCD calculations chiral symmetry should be
restored at sufficiently high temperature ($T>>T_c$).

It is, however, at the onset not at all clear that the transition
to the QGP, as it is expected to be recreated in nucleus-nucleus
collisions, proceeds between states of thermodynamic equilibrium
as sketched above. The reaction, from first contact of the
colliding nuclei to freeze-out of the created fireball, occurs on
a typical timescale of about $\unit[10]{fm/c}$ and is governed by
complex reaction dynamics so that non-equilibrium features may be
important. Likewise there can be significant rescattering of the
strongly interacting components of the system, after its
formation, that tends to obscure specific features associated with
a phase transition.

Many potential experimental signatures for the existence of the
QGP have been proposed. These can be roughly grouped into two
classes: 1) {\em evidence for bulk properties consistent with QGP
formation}, e.g. large energy density, entropy growth, plateau
behavior of the thermodynamic variables, unusual expansion and
lifetime properties of the system, presence of thermodynamic
equilibration, fluctuations of particle number or charge balance
etc, and 2) {\em evidence for modifications of specific properties
of particles thought to arise from their interactions with a QGP},
e.g. the modification of widths and masses of resonances,
modification of particle production probabilities due to color
screening (e.g. J/$\Psi$ suppression) and modification of parton
properties due to interaction with other partons in a dense medium
(e.g. jet quenching), etc.

We may ask the following questions: {\em 1) What is the
requirement for calling a state of matter a QGP}, and {\em 2) What
would constitute proof of QGP formation according to that
definition?}

As far as the first question is concerned it would seem obvious
that the the determining factor is whether the high density state
that is created in the nuclear collisions clearly has properties
that are determined by its partonic composition, beyond what is
known at the nucleon level in elementary nucleon-nucleon
collisions (e.g. p+p collisions). It has often been presupposed
that the 'plasma' should be in thermodynamical equilibrium.
However, this may not be realized within the short time scales
available for the evolution of the reaction from first contact to
freeze-out, and is perhaps not necessary in the definition of the
version of the QGP that may be observable in relativistic heavy
ion collisions. Finally, it may be asked whether chiral symmetry
restoration is essential. It would seem that even in a situation
in which the partons of the system are still (strongly)
interacting one may speak of a QGP as long as the constituents are
not restricted to individual hadrons. Thus it would appear that
{\em deconfinement} is the foremost property needed to define the
QGP state, and the one that needs to be demonstrated by
experiment.

Clearly, the observation of all, or at least of a number of the
effects listed above, in a mutually consistent fashion, would
serve to constitute a strong case for the formation of a QGP.
Ideally, the observed effects must not be simultaneously
describable within other frameworks, e.g. those based on purely
hadronic interactions and not explicitly involving the partonic
degrees of freedom. This suggests the requirement that a 'proof',
in addition to having consistency with QGP formation, also must
contain elements that are {\em only} describable in terms of QGP
formation, phase transition etc.

Finally, if a sufficiently good case exists, we may also ask if
there are any specific features that may {\em falsify} the
conclusion. To our knowledge no tests have been proposed that may
allow falsification of either a partonic scenario or a hadronic
scenario, but it would be important if any such exclusive tests
were to be formulated.

In this report we address some of the signatures discussed above,
notably the energy density, which can be deduced from the measured
particle multiplicities, the thermal and dynamical properties of
the matter at freeze-out which may be inferred from the abundances
and spectral properties of identified particles, and the
modifications of spectral properties arising from the interaction
of particles with the high energy-density medium.

\section{Reactions at RHIC: how much energy is released?}

The kinetic energy that is removed from the beam and which is
available for the production of a state such as the QGP depends on
the amount of stopping between the colliding ions.

The stopping can be estimated from the rapidity loss experienced
by the baryons in the colliding nuclei. If incoming beam baryons
have rapidity, $y_b$ relative to the CM (which has $y=0$) and
average rapidity
\begin{equation}
<y> = \int_0^{y_b} y {{dN}\over{dy}} dy  / \int_0^{y_b}
{{dN}\over{dy}} dy
\end{equation}
after the collision, the average rapidity loss is $\delta y = y_b
- <y>$~\cite{FVandOHstopping,chap3ref1}. Here dN/dy denotes the
number of net-baryons (number of baryons minus number of
antibaryons) per unit of rapidity. Thus, for the case of full
stopping: $\delta y =  y_b$.

\begin{figure}[htb]
  \begin{minipage}[t]{0.49\columnwidth}
    \includegraphics[width=\linewidth]{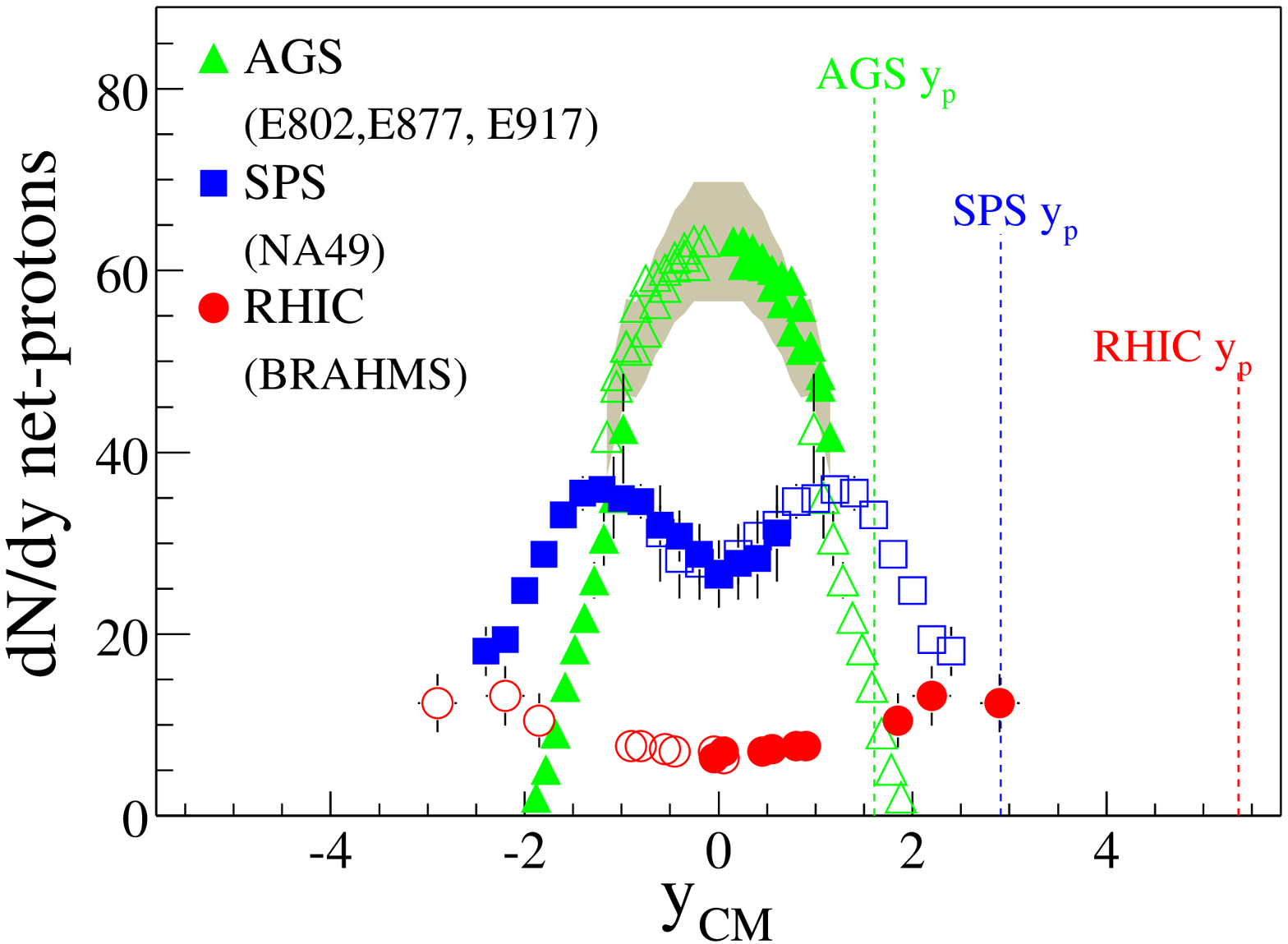}
    \caption{Rapidity density of net protons (i.e.\
      number of protons minus number of antiprotons) measured at AGS, SPS, and
      RHIC (BRAHMS) for central collisions. At RHIC, where the beam rapidity is
      $y=5.4$,
      the full distribution cannot be measured
      with current experiments, but BRAHMS will be able to extend its
      unique results to y=3.5 from the most recent high statistics Au+Au run, corresponding to
      measurements extending to 2.3 degrees with respect to the beam direction.}
    \label{fig1}
  \end{minipage}
  \hspace{0.02\columnwidth}
  \begin{minipage}[t]{0.49\columnwidth}
    \includegraphics[width=\linewidth]{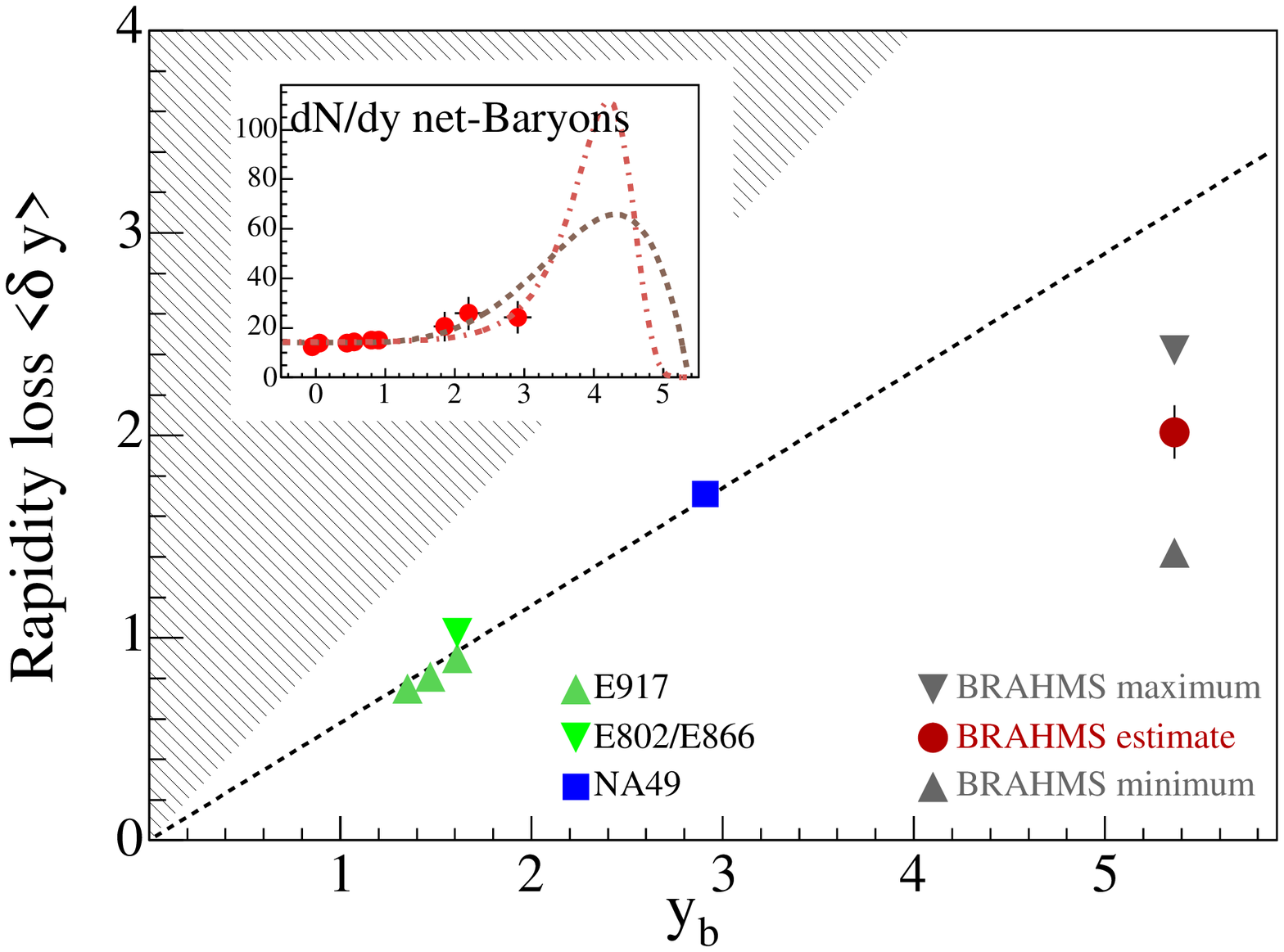}
    \caption{Insert: two possible net-baryon distributions (Gaussian in $p_T$ and 6'th
      order polynomial) respecting baryon number conservation. In going from
      net-proton to net-baryon distributions we have assumed that $N(n)\approx
      N(p)$ and have scaled hyperon yields known at midrapidity to forward
      rapidity using HIJING. Even assuming that all missing baryons are located just beyond
      the acceptance edge or at the beam rapidity, quite tight limits
      on the rapidity loss of colliding Au ions at RHIC can be set
      (main panel).}
    \label{fig2}
  \end{minipage}
\end{figure}

At AGS energies the number of produced antiprotons is quite small
and the net-baryon distribution is similar to the proton
distribution~\cite{chap3ref3,chap3ref4,chap3ref5}. The net-proton
rapidity distribution is centered around $y=0$ and is rather
narrow. The rapidity loss is about 1 for a beam rapidity of
approx. 1.6. At CERN-SPS energies ($\sqrt{s_{NN}}=\unit[17]{GeV},
\unit[158]{AGeV}$ Pb+ Pb reactions) the rapidity loss is slightly
less than 2 for a beam rapidity of 2.9~\cite{NA49-RAA}, about the
same relative rapidity loss as at the AGS.  The fact that the
rapidity loss is large on an absolute scale means, however, that
there is still a sizeable energy loss of the colliding nuclei.
This energy is available for particle production and other
excitations, transverse and longitudinal expansion. Indeed, in
collisions at the SPS, multiplicities of negatively charged
hadrons are about $dN/dy=180$ around $y=0$. At SPS another feature
is visible (see fig. 1): the net proton rapidity distribution
shows a double 'hump' with a dip around $y=0$. This shape results
from the finite rapidity loss of the colliding nuclei and the
finite width of each of the humps, which reflect the rapidity
distributions of the protons after the collisions. This picture
suggests that the reaction at the SPS is beginning to be
transparent in the sense that fewer of the original baryons are
found at midrapidity after the collisions, in contrast to the
situation at lower energies.

BRAHMS has measured the net proton rapidity distribution at RHIC
in the interval $y=0-3$ in the first run with ($0-10\%$) central
Au+Au collisions at full energy. The beam rapidity at RHIC is
about 5.4. Details of the analysis can be found in
\cite{BRAHMSnetproton}. The results are displayed in
fig.~\ref{fig1} together with the previously discussed net-proton
distributions measured at AGS and SPS. The distribution measured
at RHIC is both qualitatively and quantitatively very different
from those at lower energies indicating a significantly different
system is formed near midrapidity.

The net number of protons per unit of rapidity around $y=0$ is
only about 7 and the distribution is flat over at least the $\pm
1$ unit of rapidity. The distribution rises in the rapidity range
$y=2-3$ to an average $dN/dy \approx 12$. We have not yet
completed the measurements at the most forward angles (highest
rapidity) allowed by the geometrical setup of the experiment, but
we can exploit baryon conservation in the reactions to set limits
on the relative rapidity loss at RHIC. This is illustrated in
fig.~\ref{fig2}, which shows two possible distributions whose
integral areas correspond to the number of baryons present in the
overlap between the colliding nuclei. From such distributions one
may deduce a set of upper and lower limits for the rapidity loss
at RHIC.
Furthermore the situation is complicated by the fact that not all
baryons are measured. The limits shown in the figure includes
estimates of these effects~\cite{BRAHMSnetproton}. The conclusion
is that the {\it absolute} rapidity loss at RHIC $(\delta y =2.0
\pm 0.4)$ is not appreciably larger than at SPS. The value is
close to expectations from extrapolations of pA data at lower
energies ~\cite{chap3ref1,BuszaLedoux}. In fact the {\it relative}
rapidity loss is significantly reduced as compared to an
extrapolation of the low energy
systematics~\cite{FVandOHstopping}.

It should be noted that the rapidity loss is still significant and
that, since the overall beam energy (rapidity) is larger at RHIC
than at SPS, the {\em absolute energy loss} increases appreciably
from SPS to RHIC thus making available a significantly increased
amount of energy for particle creation in RHIC reactions.

In particular we have found that the average energy loss of the
colliding nuclei corresponds to about $\unit[73\pm 6]{GeV}$ per
nucleon\cite{BRAHMSnetproton}.
From our measurements of the particle production as a
function of rapidity (pions, kaons and protons and their
antiparticles) we can deduce not only the number of produced
particles but also their average transverse momentum and thus
their energy. Within systematic errors of both measurements we
find that the particle production is consistent with the energy
that is taken from the beam.

{\em Thus, the energy loss measurements clearly establish that as
much as $\unit[26]{TeV}$ of kinetic energy is removed from the
beam per central Au+Au collision. This energy is available for
particle production in a small volume immediately after the
collision.}

\section{Energy density}

The collision scenario that we observe at RHIC and which was
outlined in the previous section indicates that the reaction can
be viewed as quite transparent. After the collision, the matter
and energy distribution can be conceptually divided up into two
main parts, a so--called fragmentation region consisting of the
excited remnants of the colliding nuclei which have experienced an
average rapidity loss, $\delta y \approx 2$, and a central region
in which few of the original baryons are present but where
significant energy density is collected.

This picture is in qualitative agreement with the schematic one
already proposed by Bjorken 20 years ago~\cite{Bjorken83}. The
central region (an interval around midrapidity) is decoupled from
the fragments. In that theoretical scenario the energy removed
from the kinetic energy of the fragments is initially stored in a
color field strung between the receding partons that have
interacted. The linear increase of the color potential with
distance eventually leads to the production of quark-antiquark
pairs. Such pairs may be produced anywhere between the interacting
partons leading to an approximately uniform particle production as
a function of rapidity and similar spectra characteristics in each
frame of reference (boost invariance).

\begin{figure}[htb]
  \begin{minipage}[t]{0.47\columnwidth}
    \includegraphics[width=\linewidth]{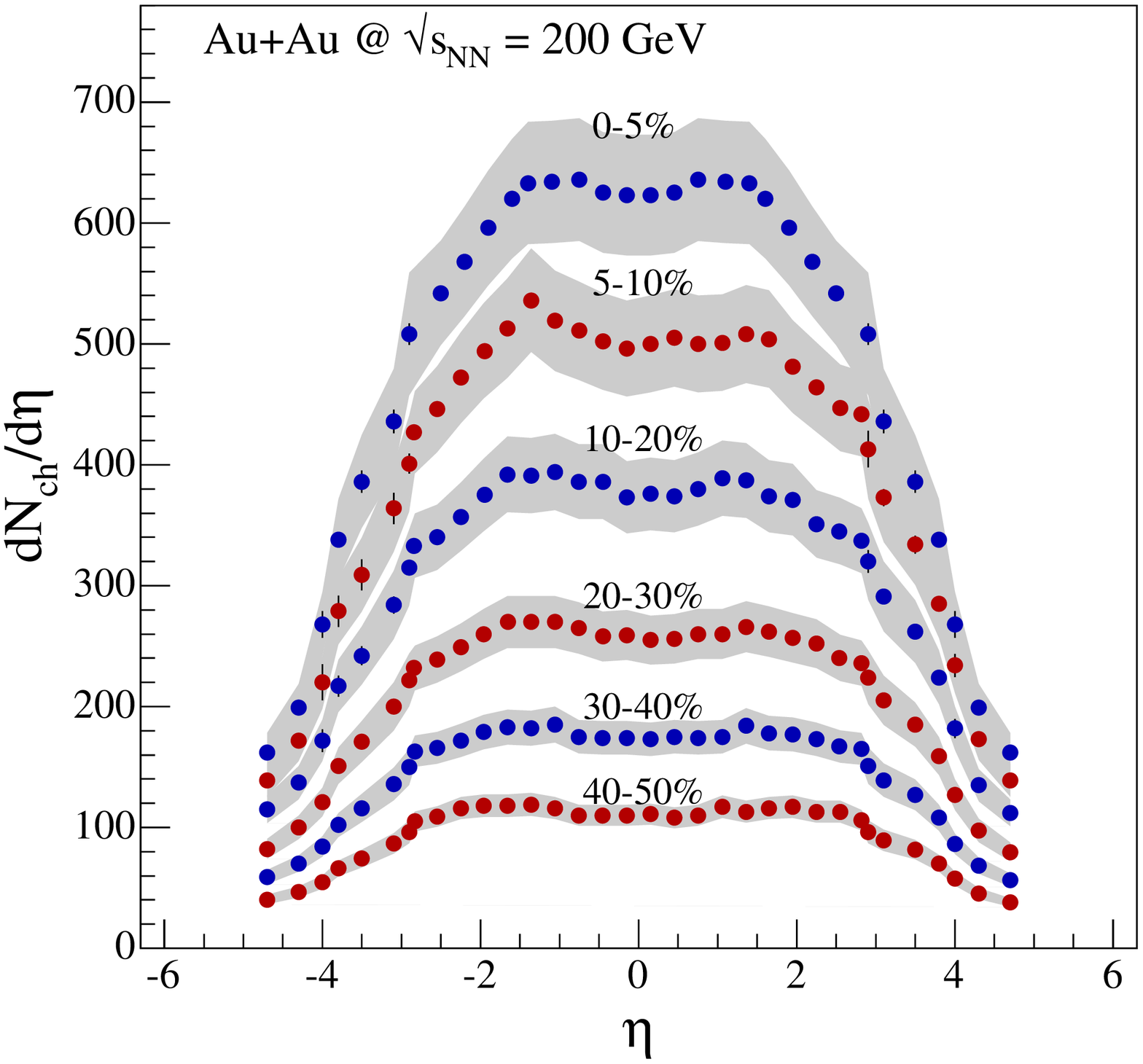}
    \caption{Pseudorapidity densities (multiplicities) of charged particles measured by BRAHMS for
      $\sqrt{s_{NN}}=\unit[200] GeV$ Au+Au collisions for various centralities.
      The integral of
      the most central distribution $0-5\%$ corresponds to about 4600
      charged particles~\cite{BRAHMSmult}.}
    \label{fig3}
  \end{minipage}
  \hspace{0.05\columnwidth}
  \begin{minipage}[t]{0.47\columnwidth}
    \includegraphics[width=\linewidth]{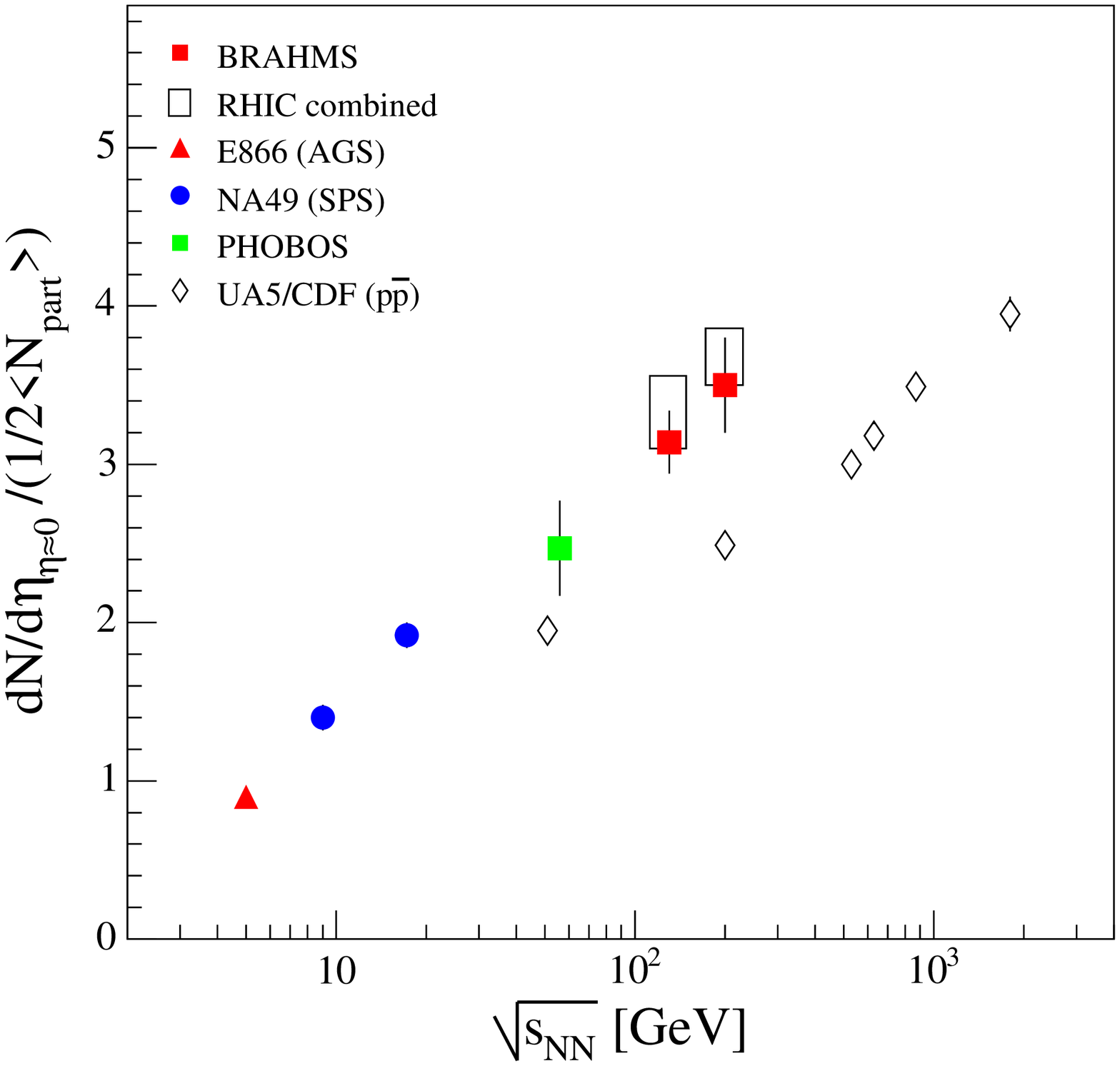}
    \caption{Multiplicity of charged particles per participant pair around midrapidity,
      as a function of $\sqrt{s_{NN}}$. The figure shows that the particle production in Au+Au
      collisions at the RHIC top energy , around $\eta=0$, exceeds that seen in p+p collisions
      by 40-50\%. }
    \label{fig:multvsenergy}
  \end{minipage}
\end{figure}

Figure~\ref{fig3} shows the overall multiplicity of charged
particles observed in Au+Au collisions at RHIC ~\cite{BRAHMSmult}
for various collision centralities and as a function of
pseudorapidity. The figure shows that the multiplicity at RHIC is
about $dN/d\eta = 625$ charged particles per unit of rapidity
around $\eta=0$ for central collisions.
Figure~\ref{fig:multvsenergy} shows that the production of charged
particles in central collisions exceeds the particle production
seen in p+p collisions at the same energy by 40-50\%, when the
yield seen in p+p collisions is multiplied by the number of
participant pairs of nucleons(participant scaling). Also we note
that the average rapidity loss in p+p collisions is $\delta y
\approx 1$. The energy available for particle production in p+p is
thus about 50\% of the beam energy, to be compared to the 73\%
found for Au+Au collisions.

Integration of the charged particle pseudorapidity distributions
corresponding to central collisions tells us that about 4600
charged particles are produced in each of the 5\% most central
collisions. Since we only measure charged particles, and not the
neutrals, we multiply this multiplicity by 3/2 to obtain the total
particle multiplicity of about 7000 particles.

From the measured spectra of pions, kaons and protons and their
antiparticles as a function of transverse momentum we can
determine the average transverse mass for each particle species
(fig.~\ref{fig:rapsystematic}). This allows us to estimate the
initial energy density from Bjorkens formula~\cite{Bjorken83}
\begin{equation}
\epsilon = {{1} \over {\pi R^2 \tau}} {{d\langle E_T \rangle}\over
{dy}}
\end{equation}
where we can make the substitution $d\langle E_T \rangle = \langle
m_T \rangle dN$ and use quantities from the measured spectral
distributions. Since we wish to calculate the energy density in
the very early stages of the collision process we may use for $R$
the radius of the overlap disk between the colliding nuclei, thus
neglecting transverse expansion. The formation time is more tricky
to determine~\cite{chap4ref3,chap4ref4}. It is often assumed to be
of the order of $\unit[1]{fm/c}$, a value that may be inferred
from the uncertainty relation and the typical relevant energy
scale (200 MeV). Under these assumptions we find that $\epsilon
\approx \unit[5]{GeV/fm^3}$, which should be considered as a lower
limit. This value of the initial energy exceeds the energy density
of a nucleus by a factor of 30 the energy density of a baryon by a
factor of 10, and the energy density for QGP formation that is
predicted by lattice QCD calculations by a factor of
5~\cite{chap4ref5,chap4ref6}.

{\em The particle multiplicities that are observed at RHIC
indicate
  that the energy density associated with particle production in the
  initial stages of the collisions largely exceeds the energy density
  of hadrons.}

\section{Is there thermodynamical and chemical equilibrium at
  RHIC?}

It has traditionally been considered crucial to determine whether
there is thermodynamical equilibration of the "fireball" in
relativistic collisions. The main reason is that, if there is
thermalization, the simple two phase model may be invoked and the
system should evidence the recognizable features of a phase
transition.

In nuclear collisions, however, the time scale available for
equilibration is very short and the entire system only lives in
the order of 10 fm/c. Consequently, it is not evident that the
system will evolve through equilibrated states. If equilibrium is
established, it would suggest that the system existed for a short
time in a state with sufficiently short mean free path. A central
issue is whether equilibrium is established in the hadronic cloud
in the later stages of the collisions just prior to freeze-out or
whether it is established on the partonic level prior to
hadronization~\cite{chap5ref2}. Thus, even if equilibration {\em per se} is
probably not a requirement for defining the QGP, it may prove to
be an important tool in {\em identifying} the QGP.

\subsection{Particle yields}

\begin{figure}[htb]
\begin{center}
 \epsfig{file=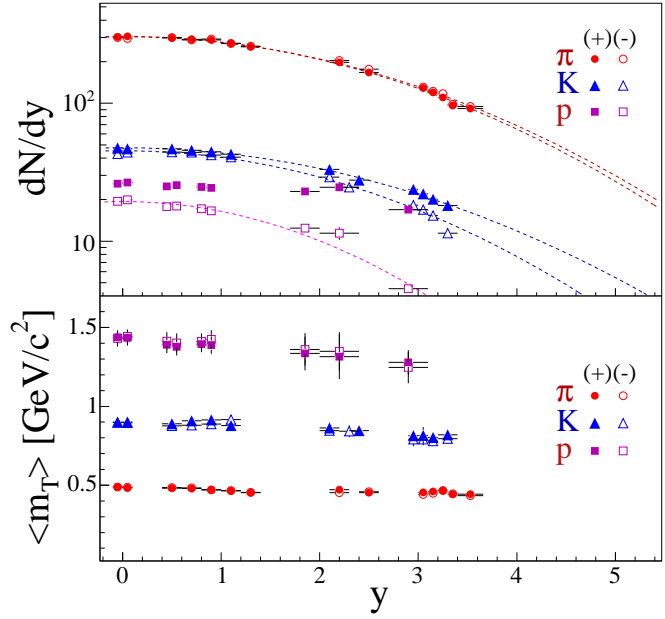,width=9cm}
  \caption{Top panel: rapidity density distribution for positive
  and negative pions, kaons and protons measured by Brahms.
  The shown data have not been corrected for feed down. The lines show Gaussian fits to the
  measured distributions.   Bottom panel: average $m_T$ distributions as
  a function of rapidity. From~\cite{DjamelPhD}.}
 \label{fig:rapsystematic}
\end{center}
\end{figure}

\begin{figure}[ht]
\begin{center}
  \epsfig{file=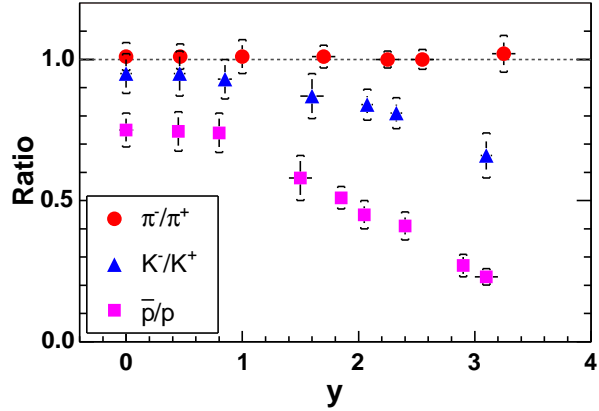,width=8cm}
  \caption{Ratios of antiparticles to particles (pions, kaons and protons)
    as a function of rapidity for $\sqrt{s_{NN}}=\unit[200]{GeV}$
    Au+Au collisions measured by the BRAHMS
    experiment~\cite{BRAHMSratios}.  For the first time in nuclear
    collisions an approximate balance between particles and
    antiparticles is seen around midapidity. Statistical and systematic errors are indicated.}
 \label{fig:antiparticleratio}
 \end{center}
\end{figure}

\begin{figure}[htb]
\begin{center}
\epsfig{file=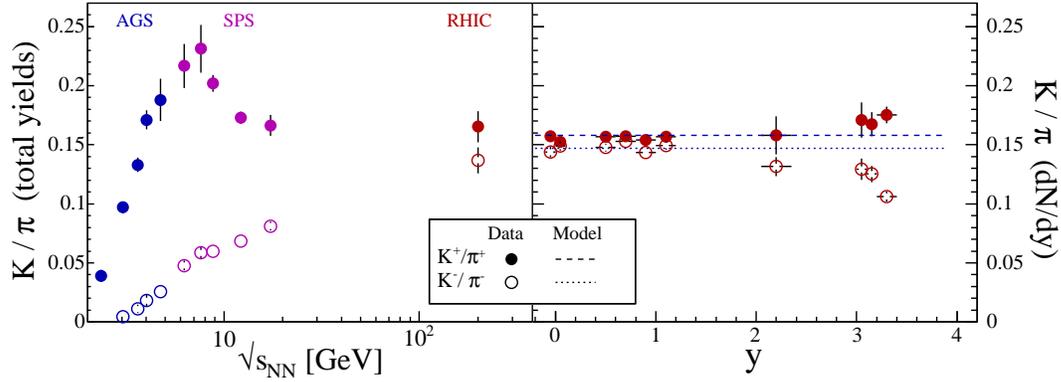,width=14cm}

  \caption{Left panel: ratios of kaons and pions of both charge signs and in the full phase space
  (i.e. integrated over azimuthal angles and over rapidity) as a function of center of
  mass energy in the nucleon-nucleon system. Right panel: the ratios at the top RHIC
  energy as a function of rapidity. At midrapidity the two ratios are about the same
  and equal to 0.15~\cite{DjamelPhD}, while at forward rapidity the ratio of  positive
  kaons and pions increases as expected for a larger baryochemical potential.
  The lines show statistical model predictions assuming a temperature of 177 MeV
  and $\mu_B = 29$ MeV.}
 \label{Koverpi}
 \end{center}
\end{figure}

\begin{figure}[htb]
\begin{center}
 \epsfig{file=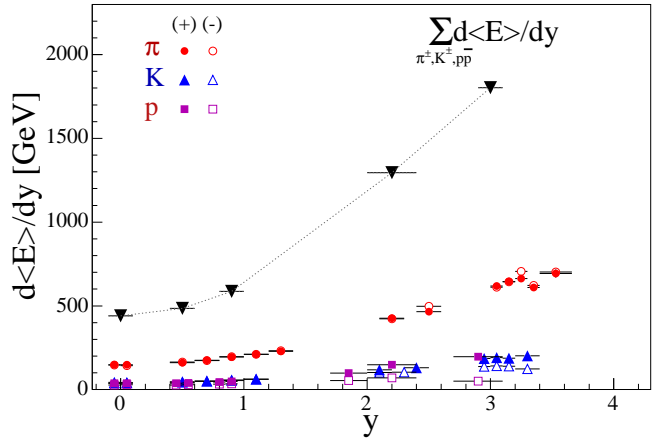,width=9cm}
  \caption{Total relativistic energy carried by charged hadrons (pions, kaons, protons and their
  antiparticles) in the rapidity interval $0 < y < 3$ deduced from the information in fig. \ref{fig:rapsystematic} and using
  the relationship $E=m_t cosh(y)$. The triangles show the sum of the individual distributions.
  Adding the expected contribution from unobserved neutral particles it can be concluded that particles
  in the range $-3<y<3$ carry about 9 TeV of total energy whereas
  particle in the range $-1<y<1$ carry about 1.5 TeV.}
 \label{fig:EvsYsystematic}
\end{center}
\end{figure}

Figure~\ref{fig:rapsystematic} shows the results of a recent and
more detailed study of particle production in central collisions
as a function of rapidity~\cite{BRAHMSnetproton,DjamelPhD}. The
figure shows the rapidity densities of pions and kaons for central
collisions. From such distributions we can construct the ratio of
the yields of particles and their antiparticles as a function of
rapidity. Figure~\ref{fig:antiparticleratio} shows the ratios of
yields of antihadrons to hadrons (posititive pions, kaons and
protons and their antiparticles). The ratio is seen to be
approaching unity in an interval of about 1.5 units of rapidity
around midrapidity, suggesting that the particle production in the
central region is predominantly from pair creation. This is true
for pions (ratio of 1), but less so for kaons (ratio=0.95) and
protons (ratio= 0.76). There are processes that break the symmetry
between particles and antiparticles that depend on the net-baryon
content discussed in the previous section. One such process that
is relevant for kaons is the associated production mechanism (e.g.
$p+p \rightarrow p + \Lambda + K^+$) which leads to an enrichment
of positive kaons in regions where there is an excess of baryons.
Support for this view is given by fig.~\ref{Koverpi}, which shows
the systematics of kaon production relative to pion production as
a function of center of mass energy. At AGS, where the net proton
density is high at midrapidity, the rapidity density of $K^+$
strongly exceeds that of $K^-$. In contrast, at RHIC, production
of $K^+$ and $K^-$ is almost equal. This situation changes,
however, at larger rapidities where the net proton density
increases.

From the measured yields of identified particles as a function of
rapidity and their momentum spectra we may calculate the total
relativistic energy carried by particles in the rapidity interval
$y=0-3$. This is shown in fig.~\ref{fig:EvsYsystematic}. By
integrating and reflecting the total energy distribution around
$y=0$ and adding the estimate contribution from neutrals we may
deduce that about 9 TeV are carried by the particles in the
rapidity range $|y|<3$.

\begin{figure}[htb]
\begin{center}
  \epsfig{file=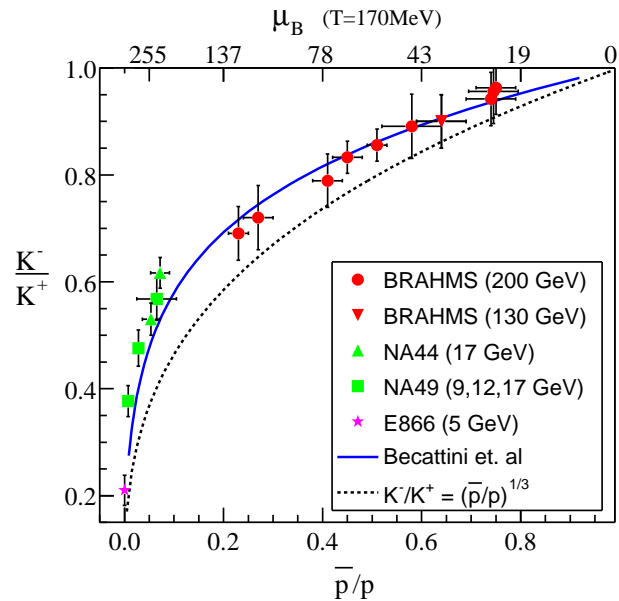,width=8cm}
  \caption{Correlation between the ratio of charged kaons and the ratio of
    antiprotons to protons. The dashed curve corresponds to equation
    4 in the text using $\mu_s =0$.  The full drawn curve is a statistical model
    calculation with a chemical freeze-out
    temperature fixed to 170 MeV~\cite{BRAHMSratios,Becattini} but
    allowing the baryochemical potential to vary. The circles denote ratios measured
    by BRAHMS at the top RHIC energy at different rapidities in the range $0<y<3$. At
    midrapidity the baryochemical potential has decreased to $\mu_B \approx 25 MeV$.}
 \label{fig:KvsPratios}
 \end{center}
\end{figure}

The particle yields measured by BRAHMS also lend themselves to an
analysis of the charged particle production in terms of the
statistical
model~\cite{BRAHMSratios,Koch86,Cleymans93,Cleymans94,PBM01statmodel,Becattini,chap5ref7}.
Figure~\ref{fig:KvsPratios} shows the ratios of negative kaons to
positive kaons as a function of the corresponding ratios of
antiprotons to protons for various rapidities at RHIC. The data
are for central collisions, and the figure also displays similar
ratios for heavy ion collisions at AGS and SPS energies. There is
a striking correlation between the RHIC/BRAHMS kaon and proton
ratios over 3 units of rapidity. Assuming that we can use
statistical arguments based on chemical and thermal equilibrium at
the quark level, the ratios can be written
\begin{equation}
{{\rho(\bar{p})}\over{\rho(p)}}= exp({{-6\mu_{u,d}}\over{T}})
\label{eq:mud}
\end{equation}
and
\begin{equation}
{{\rho(K^-)}\over{\rho(K^+)}}= exp({{-2(\mu_{u,d}
-\mu_{s})}\over{T}})
          = exp({{2\mu_s}\over{T}})\times
          [{{\rho(\bar{p})}\over{\rho(p)}}]^{{1}\over{3}}
\end{equation}

where $\rho, \mu$ and $T$ denote number density, chemical
potential and temperature, respectively. From equation \ref{eq:mud} we find
the chemical potential for $u$ and $d$ quarks to be around
$\unit[25]{MeV}$, the lowest value yet seen in nucleus-nucleus
collisions. Equation 4 tells us that for a vanishing strange quark
chemical potential we would expect a power law relation between
the two ratios with exponent 1/3. The observed correlation
deviates from the naive expectation suggesting a finite value of
the strange quark chemical potential.

A more elaborate analysis assuming a grand canonical ensemble with
charge, baryon and strangeness conservation can be carried out by
fitting these and many other particle ratios observed at RHIC in
order to obtain the chemical potentials and the temperature. It is
found that a very large collection of such particle ratios are
extremely well described by the statistical
approach~\cite{PBM01statmodel,chap5ref7}. An example of such a
procedure is shown in fig.~\ref{fig:KvsPratios} and displayed with
the full line~\cite{Becattini}. Here the temperature is
$\unit[170]{MeV}$. The point to be made is that the calculation
agrees with the data over a wide energy range (from SPS to RHIC)
and over a wide range of rapidity at RHIC. This may be an
indication that the system is in chemical equilibrium over the
considered $\sqrt{s}$ and $y$ ranges (or at least locally in the
various $y$ bins). However, that statistical fits reproduce
particle ratios is only a necessary condition for equilibration.
Separate measurements at RHIC of, for example, elliptical flow
also suggest that the system behaves collectively and thus that
the observed ratios are not just due to the filling of phase space
according to the principle of maximum entropy.

\subsection{Flow}

\begin{figure}[htb]
\begin{center}
    \includegraphics[width=8cm]{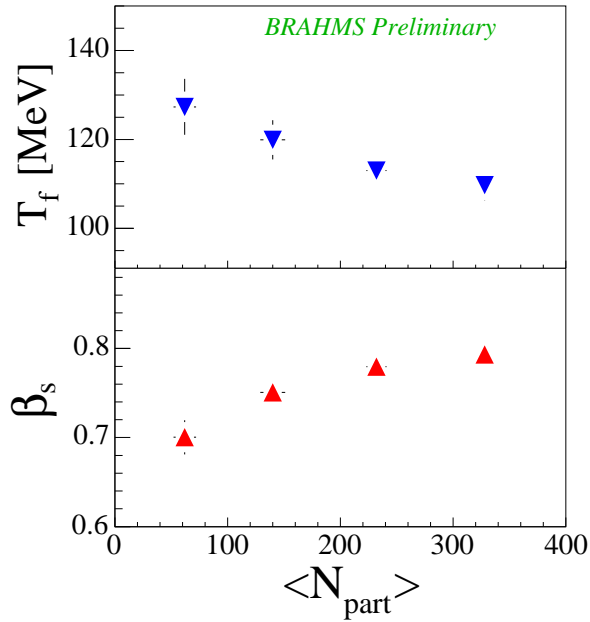}
    \caption{Temperature and (surface) transverse flow velocity at the kinetic freezeout as
      a function of collision centrality for Au+Au collisions at midrapidity. The values have
      been obtained from blastwave fits to measured transverse momentum spectra. BRAHMS
      preliminary~\cite{MMQM2004}.}
    \label{fig:kineticflow}
 \end{center}
\end{figure}

The properties of the expanding matter in the later stages of the
collisions up to the moment when interactions cease (kinetic
freeze out) can be studied from the momentum distribution of the
emitted particles. The slopes of spectra of emitted particles
depend in general on the temperature of the source from which they
were created and on kinetic effects that may alter the expected
Maxwellian distribution, such as a velocity component resulting
from an overpressure leading to an outwards flow of the matter.
This flow is expected, in the case of (at least local) thermal
equilibrium and sufficient density, to be describable by concepts
derived from fluid dynamics. One should note that the slopes of
spectra reflect the particle distributions at the time of
freeze-out when interactions have ceased.

In the so-called blastwave approach the spectrum shape is
parametrized by a function depending on the temperature and on the
transverse expansion velocity which in turn depends on the radius.
The result of such analyses for several particle/antiparticle
species indicates that the thermal (freezeout) temperature is in
the range $T=\unit[120-140]{MeV}$ and that the maximum flow
velocity is about $0.70c -0.75c$ as displayed in
Fig.~\ref{fig:kineticflow}. The first quantity is found, as
expected, to be lower than the temperature of the chemical freeze
out discussed in the previous subsection. Indeed, it would be
expected that the freeze-out of particle ratios occurs earlier
than the kinetic freeze out of the particles. The flow velocity
component is larger than what was observed at SPS energies. This
is consistent with a large pressure gradient in the transverse
direction resulting from a large initial density.
Fig.\ref{fig:kineticflow} shows results from analysis of
midrapidity particle spectra from the BRAHMS experiment using the
blastwave approach.

Another powerful tool to study the thermodynamic properties of the
source is the analysis of the azimuthal momentum distribution of
the emitted particles relative to the event plane (defined as the
direction of the impact parameter). This distribution is usually
parametrized as a series of terms depending on
$cos(n(\phi-\phi_r))$, where $\phi$ and $\phi_r$ denote the
azimuthal angles of the particle and of the reaction plane,
respectively. The coefficient ($v_1$) to the n=1 term measures the
so-called directed flow and the coefficient ($v_2$) to the n=2
term measures the elliptic flow. Elliptic flow has been analyzed
at
RHIC~\cite{chap5ref8,chap5ref9,chap5ref10,chap5ref11,chap5ref12,chap5ref13}
and has been found to reach (for many hadron species) large
($v_2$) values consistent with the hydrodynamical limit and thus
of equilibration. Model calculations
suggest~\cite{chap5ref14,chap5ref15,chap5ref16,chap5ref17,chap5ref18,chap5ref19,chap5ref20}
that the observed persistence of azimuthal momentum anisotropy
indicates that the system has reached local equilibrium very
quickly and that the equilibrium can only be established at the
{\em partonic} level when the system is very dense and has many
degrees of freedom. This explanation presupposes however that
there are many interactions and thus that the dense partonic phase
is strongly interacting.

{\em The particle ratios observed at RHIC can be well described by
concepts from statistical physics applied at the quark level, thus
assuming thermodynamical equilibrium. However this is only a
necessary condition and not a sufficient condition for
equilibration. The observation of a strong elliptic flow at RHIC
and comparison to model calculation suggests that the system is
strongly collective as must be the case for an equilibrated
system.}

\section{High $p_T$ suppression. The smoking gun of QGP?}

The discussion in the previous sections indicates that the
conditions for particle production in an interval $|y| \lesssim
1.5$ at RHIC are radically different than for reactions at lower
energies. At RHIC the central zone is baryon poor, the considered
rapidity interval appears to approximately exhibit the anticipated
boost invariant properties, the particle production is large and
dominated by pair production and the energy density appears to
exceed significantly the one required for QGP formation. The
overall scenario is therefore consistent with particle production
from a color field, formation of a QGP and subsequent
hadronization. Correlation and flow studies suggest that the
lifetime of the system is short ($<10fm/c$) and, for the first
time, there is evidence suggesting thermodynamic equilibrium
already at the partonic level.

But, is this interpretation unique? And, can more mundane
explanations based on a purely hadronic scenario be excluded? In
spite of the obvious difficulties in reconciling the high initial
energy density with hadronic volumes, a comprehensive answer to
this question requires the observation of an effect that is
directly dependent on the partonic or hadronic nature of the
formed high density zone.

\subsection{High $p_T$ suppression at midrapidity: final state partonic energy loss?}

Such an effect has recently been discovered at RHIC and is related
to the suppression of the high transverse momentum component of
hadron spectra in central Au+Au collisions as compared to scaled
momentum spectra from p+p
collisions~\cite{BRAHMShighpt,chap6ref2,chap6ref3,chap6ref4}. The
effect, originally proposed by Bjorken, Gyulassy and
others~\cite{chap6ref8,Gyulassy90,WANG92,chap6ref5} is based on
the expectation of a large energy loss of high momentum partons,
scattered in the initial stages of the collisions, in a medium
with a high density of free color charges~\cite{chap6ref9}.
According to QCD colored objects may lose energy by radiating
gluons as bremsstrahlung. Due to the color charge of the gluons,
the energy loss is proportional to the square of the length of
color medium traversed. Such a mechanism would strongly degrade
the energy of leading partons resulting in a reduced transverse
momentum of leading particles in the jets that emerge after
fragmentation into hadrons. The STAR experiment has shown that the
topology of \hpt~hadron emission is consistent with jet emission,
so that we may really speak about
jet-suppression~\cite{chap6ref10}.

\begin{figure}[ht]
\begin{center}
  \epsfig{file=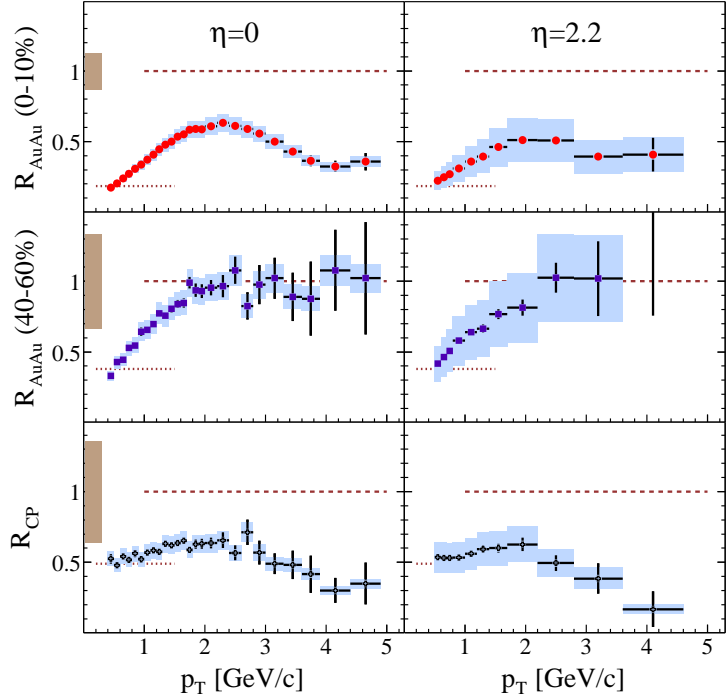,width=10cm}
  \caption{Nuclear modification factors $R_{AuAu}$ as defined in the
    text,measured by BRAHMS for central (top row) and semi-peripheral (middle row) Au+Au collisions
    at midrapidity (left) and forward pseudorapidity (right). Note the strong
    suppression of the high $p_T$ component above $p_T >2 GeV$
    seen at both rapidities. The lower row
    shows the factor $R_{cp}$, i.e. the ratio of the $R_{AuAu}$ for central and
    peripheral collisions. This ratio has the property of being independent
    of the p+p reference spectrum~\cite{BRAHMShighpt}.}
\label{fig8}
\end{center}
\end{figure}

The two upper rows of fig.~\ref{fig8} show our
measurements~\cite{BRAHMShighpt,BRAHMS-qm2002CEJ} of the
so-called nuclear modification factors for {\em unidentified}
charged hadrons from Au+Au collisions at rapidities $\eta = 0$ and
$2.2$. The nuclear modification factor is defined as:
\begin{equation}
R_{AA}=
{{d^2N^{AA}/dp_td\eta}\over{<N_{bin}>d^2N^{NN}/dp_td\eta}}.
\end{equation}
It involves a scaling of measured nucleon-nucleon transverse
momentum distributions by the number of expected incoherent binary
collisions, $N_{bin}$. In the absence of any modification
resulting from the "embedding" of elementary collisions in a
nuclear collision we expect $R_{AA}=1$ at high $p_T$. At low
$p_T$, where the particle production follows a scaling with the
number of participants, the above definition of $R_{AA}$ leads to
$R_{AA}<1$ for $p_T < \unit[2]{GeV/c}$.

In fact, it is found that $R_{AA}>1$ for $p_T>\unit[2]{GeV/c}$ in
nuclear reactions at lower energy. This enhancement, first
observed by Cronin, is associated with multiple scattering of
partons~\cite{Cronin75,chap6ref12}.

Figure~\ref{fig8} demonstrates that, surprisingly, $R_{AA}<1$ also
at high $p_T$ for central collisions at both pseudorapidities,
while $R_{AA} \approx 1$ for more peripheral collisions. It is
remarkable that the suppression observed at $p_T \approx
\unit[4]{GeV/c}$ is very large, amounting to a factor of 3 for
central Au+Au collisions as compared to $p+p$ and a factor of more
than 4 as compared to the more peripheral collisions. Such large
suppression factors are observed at both pseudorapidities.

\begin{figure}[htb]
\begin{center}
  \epsfig{file=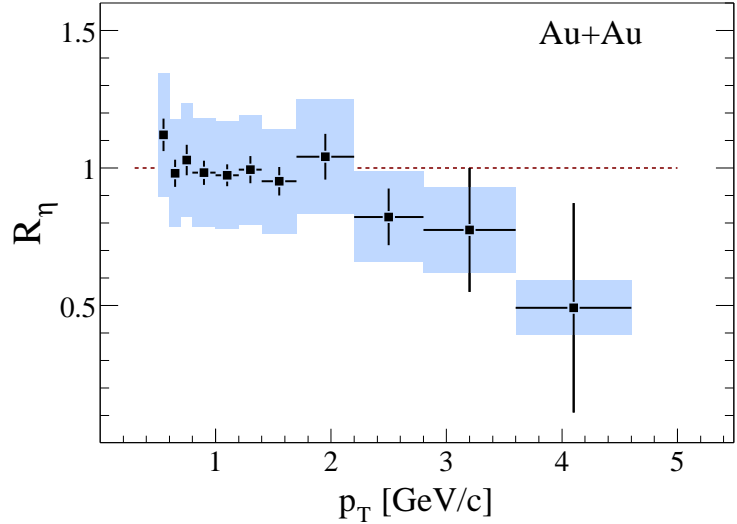,width=10cm}
  \caption{Ratio $R_\eta$ of the suppression factors $R_{cp}$ at
   pseudorapidities $\eta=0$ and $\eta=2.2$ that are shown in
   figure~\ref{fig8}. The
   figure suggest that high $p_T$ suppression persists
   (and is even more important) at forward rapidity than
   at $\eta=0$~\cite{BRAHMShighpt}.}
 \label{fig10}
 \end{center}
\end{figure}

The very large suppression observed in central Au+Au collisions
must be quantitatively understood and requires systematic
modelling of the dynamics. At $\eta =0$ the particles are emitted
at 90 degrees relative to the beam direction, while at $\eta =
2.2$ the angle is only about 12 degrees. In a naive geometrical
picture of an absorbing medium with cylindrical symmetry around
the beam direction, the large suppression seen at forward angles
suggests that the suppressing medium is extended also in the
longitudinal direction. Since the observed \hpt~suppression is
similar or even larger at forward rapidity as compared to
midrapidity (see fig.~\ref{fig10}) one might be tempted to infer a
longitudinal extent of the dense medium which is approximately
similar to its transverse dimensions. However, the problem is more
complicated, due to the significant transverse and in particular
longitudinal expansion that occurs as the leading parton
propagates through the medium, effectively reducing the densities
of color charges seen. Also other \hpt~suppressing mechanisms may
come into play at forward rapidities (see discussion on the Color
Glass Condensate in the following chapter).

\begin{figure}[htb]
\begin{center}
  \epsfig{file=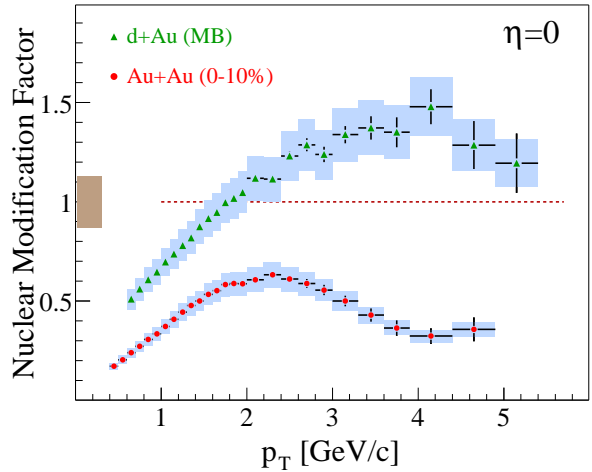,width=8cm}
  \caption{Nuclear modification factors measured for central Au+Au collisions
    and minimum bias d+Au collisions at
    $\sqrt{s_{NN}}=\unit[200]{GeV}$, evidencing the important
    \hpt~suppression observed in central Au+Au collisions~\cite{BRAHMShighpt} which
    is absent in the d+Au reactions. The shaded band around the points indicates the
    systematic errors. The shaded box on the ordinate around unity shows the estimated
    uncertainty on the value of $N_{bin}$.  }
 \label{fig9}
\end{center}
\end{figure}

It has been conjectured that the observed \hpt~suppression might
be the result of an entrance channel effect, for example as might
arise from a limitation of the phase space available for parton
collisions related to saturation effects \cite{gluonsat} in the
gluon distributions inside the swiftly moving colliding nucleons
(which have $\gamma = 100$). As a test of these ideas we have
determined the nuclear modification factor for d+Au minimum bias
collisions at $\sqrt{s_{NN}}=\unit[200]{GeV}$. The resulting
$R_{dAu}$ is shown in fig.~\ref{fig9} where it is also compared to
the $R_{AuAu}$ for central collisions previously shown in
fig.~\ref{fig8}. No high $p_T$ jet suppression is observed for
d+Au \cite{BRAHMShighpt,chap6ref13,PHOBOSptAuAu,chap6ref15}. The
$R_{dAu}$ distribution at $y=0$ shows a Cronin enhancement similar
to that observed at lower
energies~\cite{NA49-RAA,WA98-RAA,CERES-RAA}. At $p_T \approx
\unit[4]{GeV/c}$ we find a ratio $R_{dAu}/R_{AuAu} \approx 4-5$.
These observations are consistent with the smaller transverse
dimensions of the overlap disk between the d and the Au nuclei and
also appear to rule out initial state effects.

High $p_T$ suppression at forward rapidities may also be expected
to arise from the possible Color Glass Condensate phase in the
colliding nuclei (see the discussion in the next section). There
is little doubt that systematic studies of the high $p_T$ jet
energy loss as a function of the thickness of the absorbing medium
obtained by varying the angle of observation of high $p_T$ jets
relative to the event plane and the direction of the beams will be
required in order to understand in detail the properties of the
dense medium.

\subsection{The flavor composition}

\begin{figure}[htb]
 \begin{center}
   \includegraphics[width=0.8\linewidth]{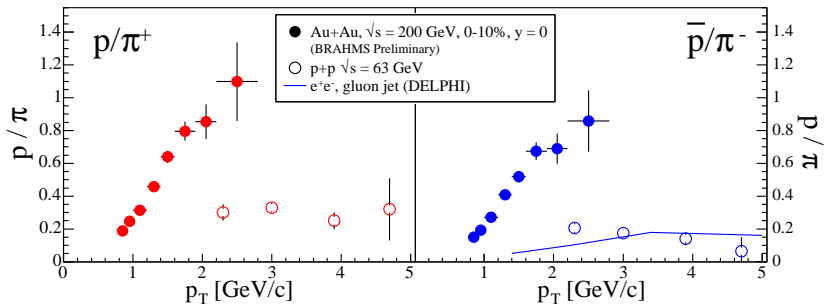}
   \caption{ Ratios of particle yields $p/\pi^+$ (left) and $\bar{p}/\pi^-$ (right) measured at mid-rapidity
     for 0-10\% central Au+Au collisions at $\sqrt{s_{NN}} =
     \unit[200]{GeV}$.  The error bars show the statistical errors.
     The systematic errors are estimated to be smaller than 8\%. Data at
     $\sqrt{s} = \unit[63]{GeV}$ for $p+p$ collisions~\cite{chap6ref18} are
     also shown (open circles).  The solid line in the right hand panel
     is the $(p+\bar{p})/(\pi^++\pi^-)$ ratio measured for gluon
     jets~\cite{chap6ref19} in $e^+ + e^-$ collisions.}
 \label{fig:pPiRatioRap0}
\end{center}
\end{figure}

With its excellent particle identification capabilities BRAHMS can
also study the dependence of the high $p_T$ suppression on the
type of particle. Preliminary
results~\cite{BRAHMS-qm2002CEJ,chap6ref17} indicate that mesons
(pions and kaons) experience high $p_T$ suppression while baryons
(protons) do not. The reason for this difference is at present not
well understood.

\begin{figure}[htbp]
\begin{center}
  \epsfig{file=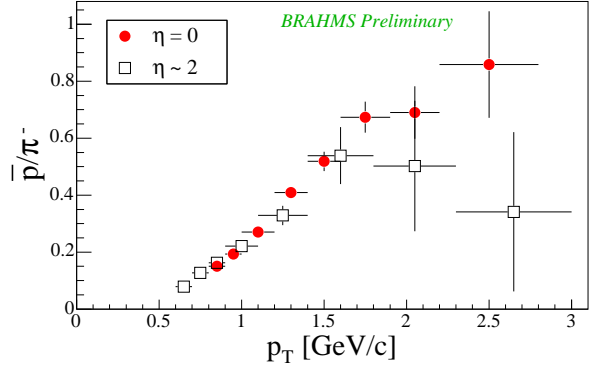, width=8.0cm}
  \caption{Comparison of the ratios yields of $\bar{p}/\pi^-$ at rapidities $y=0$ and
  $y=2.2$. In spite of small statistics the data suggest that a forward rapidity the
  flow may be weaker resulting in a derecreased yield of antiprotons relative to pions above
  $p_T \approx 2 GeV$. BRAHMS preliminary~\cite{chap6ref17}}
 \label{fig:pPiRatioModelComparison}
\end{center}
\end{figure}

The observed differences may be a consequence of baryons being
more sensitive to flow, because of their larger mass, than mesons.
The flow contribution leads a flatter transverse momentum spectrum
for baryons than for mesons, thus possibly compensating for a high
$p_T$ suppression effect similar to that of the mesons. It is also
possible that the difference reflects details associated with the
fragmentation mechanism that leads to different degrees of
suppression of the high $p_T$ component for 2 and 3 valence quark
systems. Finally the difference may reflect the mechanism of
recombination for 3 quarks relative to that for 2 quarks in a
medium with a high density of quarks.

Figure~\ref{fig:pPiRatioRap0} shows a recent investigation by
BRAHMS (ref. \cite{chap6ref17}) of the baryon to meson ratios at
mid-rapidity $p/\pi^+$ and $\bar{p}/\pi^-$, as a function of $p_T$
for the 0-10\% most central Au+Au collisions at $\sqrt{s_{NN}} =
200$ GeV. The ratios increase rapidly at low $p_T$ and the yields
of both protons and anti-protons are comparable to the pion yields
for $p_T > \unit[2]{GeV/c}$. The corresponding ratios for $p_T >
\unit[2]{GeV/c}$ observed in $p+p$ collisions at $\sqrt{s} = 62$
GeV~\cite{chap6ref18} and in gluon jets produced in $e^++e^-$
collisions~\cite{chap6ref19} are also shown. The increase of the
$p/\pi^+$ and $\bar{p}/\pi^-$ ratios at high $p_T$, seen in
central Au+Au collisions,  relative to the level seen in $p+p$ and
$e^++e^-$ indicates significant differences in the overall
description, either at the production or fragmentation level.

Figure~\ref{fig:pPiRatioModelComparison} shows the comparison of
BRAHMS data for the ratio of antiprotons to negative pions at
$\eta = 0$ and $2.2$. Although statistics at high transverse
momentum are low there are indications that the ratio is smaller
at the higher rapidity for $p_T > 2 GeV$. Recent calculations
based on a parton recombination
scenario~\cite{chap6ref20,chap6ref21,chap6ref22} with flow at the
partonic level appear to be able to describe the data at
midrapidity, while calculations omitting flow fall short of the
data already at $p_T \approx \unit[1.5]{GeV}$.

The experimental and theoretical investigation of these questions
is, however, still in its infancy. These issues can and will be
addressed in depth through the analysis of the large data set
collected by BRAHMS in the high luminosity Au+Au run of year 2004.

\subsection{High $p_T$ suppression at lower energy?}

\begin{figure}[htb]
\begin{center}
  \epsfig{file=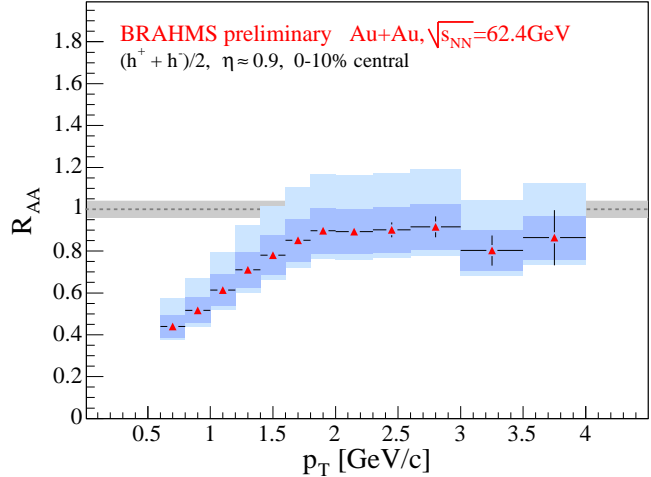,width=9cm}
  \caption{Nuclear modification factor $R_{AuAu}$ measured by BRAHMS
  for charged hadrons at $\eta=0.95$ for $0-10\%$ central
  Au+Au collisions at $\sqrt{s_{nn}}= 62.4 GeV$~\cite{ekmanthesis}. The
  dark shaded band indicates
  the systematic errors on the data, the lighter shaded band the combined
  estimated systematic error
  on the Au+Au data data and the p+p reference. }
 \label{fig:62GeV}
 \end{center}
\end{figure}

The short commissioning run for Au+Au collisions at
$\sqrt{s_{NN}}=\unit[62.4]{GeV}$ has allowed us to carry out a
first analysis of the high $p_T$ suppression of charged hadrons at
an energy of about 1/3 the maximum RHIC energy and about 3.5 times
the maximum SPS energy. Preliminary results are shown in figure
\ref{fig:62GeV} for nuclear modification factor calculated for the
sum of all charged hadrons measured at 45 degrees($\eta=0.9$) with
respect to the beam direction. The data have been compared to
reference spectra measured in $\sqrt{s_{NN}}=\unit[63]{GeV}$ p+p
collisions at the CERN-ISR. The figure shows that the high $p_T$
data are less suppressed at $\sqrt{s_{NN}}=\unit[62.4]{GeV}$ than
at $\sqrt{s_{NN}}=\unit[200]{GeV}$. This is consistent with recent
results from PHOBOS \cite{chap6ref23}. For comparison, at SPS
energies no high $p_T$ suppression was observed, (albeit a
discussion has surfaced regarding the accuracy of the reference
spectra at that energy. It thus seems the suppression increases
smoothly with energy.

{\em The remarkable suppression of high $p_T$ jets at mid-rapidity
seen at RHIC is an important signal that evidences the interaction
of particles originating from hard parton scatterings with the
high energy density medium created in the collisions. The
quantitative understanding of the observed high $p_T$ suppression,
as a function of energy, should be able to determine whether this
interaction is at the partonic or hadronic level. This needs to be
supplemented by detailed studies of the flavor dependence of the
suppression mechanism.}

\section{The color glass condensate: a model for the initial state of nuclei?}

As part as the study of the high $p_T$ suppression in
nucleus-nucleus collisions BRAHMS has investigated the rapidity
dependence of the nuclear modification factors as a function of
rapidity ($\eta = 0,1, 2.2, 3.2)$ in d+Au collisions at
$\sqrt{s_{NN}}=\unit[200]{GeV}$. As discussed in the previous
section the measured nuclear modification factors for d+Au are
consistent with the absence of high $p_T$ suppression around
midrapidity. This may be taken as direct evidence for the fact
that the strong high $p_T$ suppression seen in Au+Au collisions
around $y=0$ is not due to particular conditions of the colliding
nuclei (initial state effects)
\cite{chap6ref13,PHOBOSptAuAu,chap6ref13} and \cite{BRAHMShighpt}.

\begin{figure}[htb]
\begin{center}
  \epsfig{file=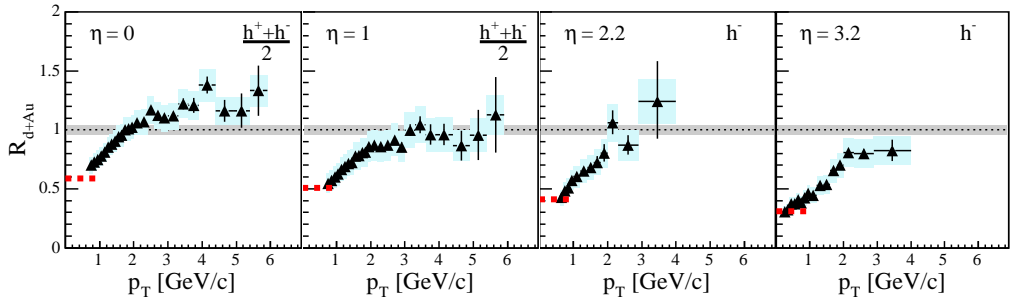,width=15cm}
  \caption{Evolution of the nuclear modification factors measured
  by BRAHMS for the 10\% most central d+Au collisions
  at $\sqrt{s_{NN}}=\unit[200]{GeV}$, as a function of pseudorapidity $\eta$
  ~\cite{BRAHMSnmf}.}
 \label{fig:RdAu-CGC}
 \end{center}
\end{figure}

At forward rapidity in d+Au collisions, however, BRAHMS has
observed \cite{BRAHMSnmf} a marked high $p_T$ suppression starting
already at $\eta=1$ (see Fig.~\ref{fig:RdAu-CGC}) and increasing
smoothly in importance with increasing pseudorapidity (up to
$\eta=3.2$). It has been proposed that this effect at forward
rapidity~\cite{KLM} is related to the initial conditions of the
colliding d and Au nuclei, in particular to the possible existence
of the Color Glass Condensate (CGC).

The CGC is a description of the ground state of swiftly moving
nuclei prior to collisions~\cite{GandMcL_25_30}. Due to the non
Abelian nature of QCD, gluons self interact which results in
nuclei containing a large number of low--$x$ gluons ($x$ is the
fraction of the longitudinal momentum carried by the parton) that
appears to diverge (grow) with decreasing $x$. There is however, a
characteristic momentum scale, termed the saturation scale, below
which the gluon density saturates. This effect sets in when x
becomes small and the associated  gluon wave length
($\frac{1}{m_{p}x}$) increases to nuclear dimensions. In such a
regime gluons may interact and form a coherent state reminiscent
of a Bose-Einstein condensate. Early indications for the formation
of such non-linear QCD systems have been found in lepton-hadron or
lepton-nucleus collisions at HERA \cite{HERAdata} and have been
described by the so called ``Geometric Scaling'' model
\cite{Golec}.

The density of gluons $\frac{dN_{g}}{d(ln(1/x))} \sim
\frac{1}{\alpha_{s}}$ in such a saturated system is high, since
$\alpha_{s}$, the strong interaction running coupling constant,
decreases as the energy increases. The system can therefore be
described as a (semi)classical field, and techniques borrowed from
field theory can be employed to find the functional form of the
parton distributions in the initial state \cite{McLerranVenu}.

Saturation in the wave function sets in for gluons with transverse
momentum $Q^2 < Q^{2}_{s} =
A^{\frac{1}{3}}(\frac{x_{0}}{x})^{\lambda} \sim
A^{\frac{1}{3}}e^{\lambda y}$. A value of $\lambda \sim 0.3$ is
estimated from fits to HERA data \cite{Lambda}. The dependence of
the saturation scale $Q_{s}$ on the atomic number of the target
and rapidity suggests that saturation effects can be studied with
heavy nuclei at large rapidities.

Collisions between heavy ions with energies $E=\unit[100]{AGeV}$
may therefore provide a window to the study of low--$x$ gluon
distributions of swiftly moving nuclei. In particular, head-on
collisions between deuterons and gold nuclei in which hadrons,
produced mostly in gluon-gluon collisions, are detected, close to
the beam direction but away from the direction of motion of the
gold nuclei, allow the low--$x$ components of the wave function of
the gold nuclei to be probed.

\begin{figure}[htb]
\begin{center}
  \epsfig{file=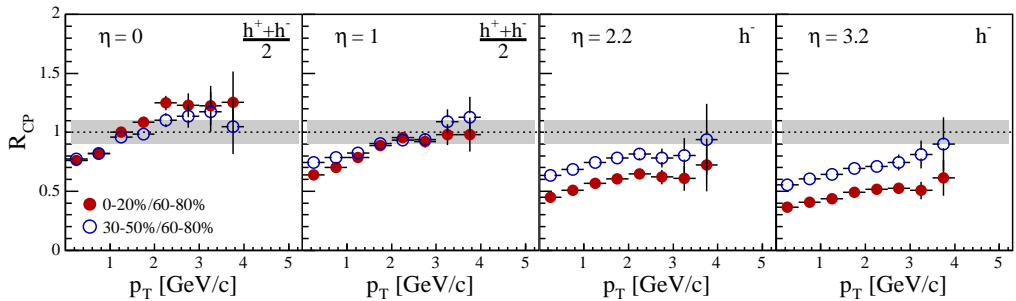,width=15cm}
  \caption{Central to peripheral ratios $R_{CP}$ as a function
  of pseudorapidity measured by BRAHMS
  for d+Au collisions at the RHIC top energy~\cite{BRAHMSnmf}. The filled circles represent
  the central-to-peripheral (0-20\% over 60-80\%) ratio. The open circles the semicentral-to-peripheral
  (30-50\% over 60-80\%) ratio. The shaded band around unity
  indicates the uncertainty associated with the values of the number
  of binary collisions at the different centralities.}
 \label{fig:RCP-CGC}
 \end{center}
\end{figure}

The centrality dependence of the nuclear modification factors
provides additional information on the mechanism underlying the
observed suppression. Fig. \ref{fig:RCP-CGC} shows the $R_{CP}$
factors, defined as the ratios of the nuclear spectra for central
(0-20\%) and peripheral (60-80\%) collisions (closed points) and
for semicentral (30-50\%) and peripheral collisions(open points),
suitably scaled by the corresponding number of binary collisions,
versus $p_T$ and $\eta$. There is a substantial change in $R_{CP}$
as a function of $\eta$. At $\eta=0$ the central-to-peripheral
collisions ratio is larger than the semicentral-to-peripheral
ratios suggesting an increased Cronin type multiple scattering
effect in the more violent collisions. In contrast, the ratio of
the most central collisions relative to the peripheral, as
compared to the semicentral-to-peripheral, is the most suppressed
at forward rapidities, suggesting a suppression mechanism that
scales with the centrality of the collisions.

The observed suppression of yields in d+Au collisions (as compared
to p+p collisions) has been qualitatively predicted by various
authors{~\cite{Dumitru,Jalilian-Marian,Wiedemann,KKT}, within the
Color Glass Condensate scenario. Recently, a more quantitative
calculation has been carried out \cite{tuchin04} which compares
well with the data. Other authors~\cite{Vitev,XWang} have
estimated the nuclear modification factors based on a two
component model that includes a parametrization of perturbative
QCD and string breaking as a mechanism to account for soft
coherent particle production using HIJING. HIJING uses the
mechanism of gluon shadowing to reduce the number of gluon-gluon
collisions and hence the multiplicity of charged particles a lower
$p_t$. HIJING has been shown to give a good description of the
overall charged particle multiplicity in d+Au collisions. Vogt has
used realistic parton distribution functions and parametrizations
of nuclear shadowing to give a reasonable description of the
minimum bais data though not of the centrality
dependence~\cite{VogtRda}.

The high $p_T$-suppression in Au+Au collisions at large rapidites
discussed earlier suggests that there may be two competing
mechanisms responsible for the observed high $p_T$ suppression in
energetic Au+Au collisions, each active in its particular rapidity
window. It has been proposed \cite{intro_GandMcL} that the high
$p_T$ suppression observed around midrapidity reflects the
presence of an incoherent (high temperature) state of quarks and
gluons while the the high $p_T$ suppression observed at forward
rapidities bears evidence of a dense coherent partonic state.
Clearly, additional analysis of recent high statistics data for Au
+Au collisions at high rapidites, as well as firmer theoretical
predictions are needed to understand the quantitative role of
gluon saturation effects in energetic nucleus-nucleus collisions.

{\em The suppression of high $p_T$ particles seen at forward
rapidities in nucleus-nucleus collisions is a novel and unexpected
effect and may be related to a new collective partonic state that
describes nuclei at small~x, and hence the initial conditions for
in energetic nucleus-nucleus collisions.}

\section{Conclusions and perspectives}

The results from the first round of RHIC experiments clearly show
that studies of high energy nucleus-nucleus collisions have moved
to a qualitatively new physics domain characterized by a high
degree of reaction transparency leading to the formation of a near
baryon free central region. There is appreciable energy loss of
the colliding nuclei, so the conditions for the formation of a
very high energy density zone with approximate balance between
matter and antimatter, in an interval of $|y| \lesssim 1.5 $
around midrapidity are present.

The indications are that the initial energy density is
considerably larger than $ \unit[5]GeV/fm^3$, i.e. well above the
energy density at which it is difficult to conceive of hadrons as
isolated and well defined entities. Analysis within the framework
of the statistical model of the relative abundances of many
different particles containing the three lightest quark flavors
suggest chemical equilibrium at a temperature in the vicinity of
$T=\unit[175]MeV$ and a near-zero light quark chemical potential.
This temperature compares  well with the prediction of lattice QCD
calculations. The conditions necessary for the formation of a
deconfined system of quarks and gluons therefore appear to be
present.

However, there are a number of features, early on considered as
defining the concept of the QGP, that do not appear to be realized
in the current reactions, or at least have not (yet?) been
identified in experiment. These are associated with the
expectations that a QGP would be characterized by a vanishing
interaction betweens quarks and exhibit the features of chiral
symmetry restoration and, furthermore, that the system would
exhibit a clear phase transition behavior. Likewise, it was
originally expected that a QGP phase created in nuclear collisions
would be characterized by a long lifetime (up to 100 fm/c) and by
the existence of a mixed phase exhibiting large fluctuations of
characteristic parameters. In contrast, the present body of
measurements compared to theory suggest a short lifetime of the
system, a large outward pressure, and significant interactions
most likely at the parton level that result in a (seemingly)
equilibrated system with fluid-like properties. Thus, the high
density phase that is observed, is not identical to the the nearly
ideal QGP as it was imagined a decade or two ago.

However, the central question is whether the properties of the
matter as it is created in todays high energy nucleus--nucleus
collisions clearly bears the imprint of a system characterized by
quark and gluon degrees of freedom over a range larger that the
characteristic dimensions of the nucleon. We know that in nuclei
the strong interaction is mediated by a color neutral objects
(mesons). Is there experimental evidence that clearly demonstrates
interactions based on the exchange of objects with color over
distances larger than those of conventional confined objects?

The best candidate for such an effect is clearly the suppression
of high transverse momentum particles observed in central Au+Au
collisions by the four experiments at RHIC. The remarkably large
effect that is observed (a suppression by a factor of 3-5 as
compared to peripheral and d+Au collisions) appears readily
explainable by radiation losses due to the interaction of high
$p_T$ partons with an extended medium (of transverse dimensions
considerably larger than nucleon dimensions) consisting of
deconfined color charges. Current theoretical investigations,
which recently have progressed to attempt first unified
descriptions of the reaction evolution, indicate that scenarios
based on interactions between hadronic objects cannot reproduce
the magnitude of the observed effect.

The interpretation of current data relies heavily on theoretical
input and modelling, in particular on the apparent necessity to
include partonic degrees of freedom in order to arrive at a
consistent description of many of the phenomena observed in the
experimental data. Seen from a purely experimental point of view
this situation is somewhat unsatisfying, but probably not
unexpected, nor avoidable, considering the complexity of the
reaction and associated processes.

It is also clear that the unravelling of the physics of the matter
state(s) observed at RHIC has just begun. In spite of the
impressive advances that have been made in the last three years
there are still many issues to be understood in detail, such as
the differences in the \hpt~suppression of baryons and mesons and
the quantitative energy and rapidity dependence of the final and
initial state \hpt~suppression. Undoubtedly future measurements
will shed new light on these and many other questions. We should
not forget, however, that there are also significant challenges
for theory. In the opening chapters of this document we remarked
on the requirement that scientific paradigms must be falsifiable.
We have yet to see a fully self consistent calculation of the
entire reaction evolution at RHIC that in an unambiguous way
demonstrates the impossibility of a hadronic description.

In conclusion, we find that the body of information obtained by
BRAHMS and the other RHIC experiments in conjunction with the
available theoretical studies is strongly suggestive of a high
density system that cannot be characterized solely by hadronic
degrees of freedom but requires a partonic description.
Indications are that such a partonic state is not characterized by
vanishing interaction of its constituents, but rather by a
relatively high degree of coherence such as the one characterizing
fluids. At the same time intriguing suggestions of a coherent
partonic state at low x in the colliding nuclei has been found.

{\em There is no doubt that the experiments at RHIC have revealed
a plethora of new phenomena that for the most part have come as a
surprise. In this sense it is clear that the matter that is
created at RHIC differs from anything that has been seen before.
What name to give it must await our deeper understanding of this
matter.}

\section{Acknowledgements}

This work was supported by the Division of Nuclear Physics of the
Office of Science of the U.S. Department of Energy under contracts
DE-AC02-98-CH10886, DE-FG03-93-ER40773, DE-FG03-96-ER40981, and
DE-FG02-99-ER41121, the Danish Natural Science Research Council,
the Research Council of Norway, the Jagiellonian University
Grants, the Korea Research Foundation Grant, and the Romanian
Ministry of Education and Research (5003/1999, 6077/2000). We
thank the staff of the Collider-Accelerator Division at BNL for
their excellent and dedicated work to deploy RHIC and their
support to the experiment.

\end{document}